# Why LP cannot solve large instances of NP-complete problems in polynomial time

Radosław Hofman, *Poznań 2006*

*Abstract*—This article discusses ability of Linear Programming models to be used as solvers of NP-complete problems. Integer Linear Programming is known as NP-complete problem, but non-integer Linear Programming problems can be solved in polynomial time, what places them in P class.

During past three years there appeared some articles using LP to solve NP-complete problems. This methods use large number of variables ($O(n^9)$) solving correctly almost all instances that can be solved in reasonable time. Can they solve infinitively large instances? This article gives answer to this question

## I. INTRODUCTION

Unknown relation between P and NP [3] complexity classes remains to be one of significant non solved problems in complexity theory. P complexity class consists of problems solvable by Deterministic Turing Machine (DTM) in polynomially bounded time, while NP complexity class consists of problem solvable by Non Deterministic Turing Machine (NDTM) in polynomially bounded time. This means that DTM can verify solution of every NP problem in polynomially bounded time even if polynomial algorithm for finding this solution is unknown [12].

Significant subclass of NP problems is known as NP-complete class. Problems from this class have ability to represent any other problems from whole NP complexity class. In 1970 S. Cook presented in [2] first reduction from any NP problem to Boolean Satisfiability Problem, and two years after R. Karp proved that 21 other problems are in NP-complete class showing many-one polynomial time reductions to these problems [10]. If then anyone shows algorithm solving any NP-complete problem in polynomially bounded time then any of NP problems may be solved in no more then $O(n^c)$ steps, where *n* stands for instance size and *c* is some constant value [12].

Recently there appeared some publications presenting usage of Linear Programming for solving famous NP-complete problems known as Traveling Salesman Problem (TSP) [4], [7], Quadratic Assignment Problem (QAP) [5] and other. This article summarizes counter example for one of these publications [4] with comment to general approach of using Linear Programming to find solution for NP-complete problems.

## II. REPRESENTATION

### A. Modeling

Storing real world or mathematical objects on every computing machine requires some model to present them. For example numbers are stored as binary streams. Complex objects may be stored in simplified form if there is a need of space or time savings. It is important to notice that sometimes this simplification may cause some loss of precision resulting in differences in outputs from model examination and expected results (if examination was performed not on model but on original object).

Let us consider as an example examination if given function $f(x)$ is monotonic. Our model would store values of function for every integer value of *x*. Checking if function is monotonic requires comparison between $f(x-1) \leq f(x)$. Let us now consider function $f(x)=\sin(2 \cdot x \cdot \pi)+x$. Mentioned model would have considered this function as monotonic while it is not.

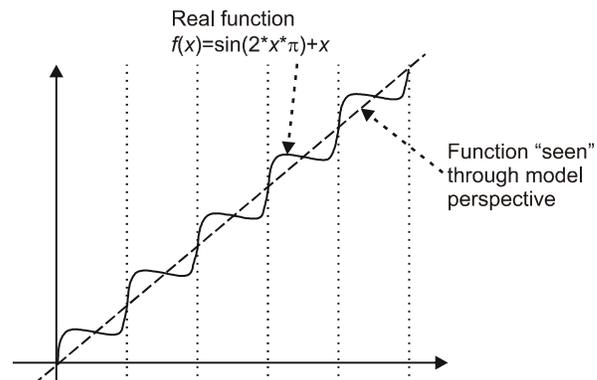

Figure 1 Difference between model and real function

Last important property is discreteness of data stored in computing machines memory. Even decimals are stored on some limited space, what means that they can represent no more then $2^b$ different values, where *b* represent number of bites used to store number.

### B. Single solution

Combinatorial problems from NP class have common property – their solution may be verified in polynomially bounded time, what means that it also can be stored in polynomially bounded space (verifying must begin with "reading" of solution).

We may see that minimum space required to store solution is $\Omega(\log(k))$, where $k$ is number of possible solutions. If then problem had $2^n$ possible solutions each may be stored on $\Omega(n)$ bytes, and for $n!$ possible solutions each may be stored on $\Omega(n \log(n))$ symbols.

### C. Set of solutions

Storing set of solutions for problems from NP class is much more expensive in terms of space used. We have to store:
1) list of solutions, or
2) set of solutions identification.

Storing list of $m$ solutions is very simple – it is list of single solutions so it requires $\Omega(m*n)$. The problem is when number of solutions to be stored refer to number of possible solutions. If one would like to store $1/d$ of all solutions then list representation requires $\Omega(2^n*n / d)$ what means it is $\Omega(2^n)$.

Second way requires to point out object being set from power-set over $2^n$. This means that there are possible $O\left(2^{2^n}\right)$ objects, and pointing one of them requires $\Omega\left(\log\left(2^{2^n}\right)\right) = \Omega(2^n)$. For possible $n!$ solutions lower bound estimation is also exponential.

No matter which type of representation was chosen, in general storing set of solutions requires $\Omega(2^n)$ symbols.

## III. LINEAR PROGRAMMING FOR COMBINATORIAL PROBLEMS

### A. Polytope

Combinatorial problem instances may be considered as kind of polytopes as described in [1], [6], [13]. We have to observe one property of these polytopes independent of number of dimensions they exist in: number of points polytope consists of is equal to number of possible solutions to problem. In [13] there is proposed succinct form of storing polytope only when variables are assigned only with values 0 or 1.

Every $n$-th dimensional polytope has $n$-1 dimensional facets representing this polytope in lower number of dimensions. Number of facets is in general exponential for polytopes with exponential number of vertexes.

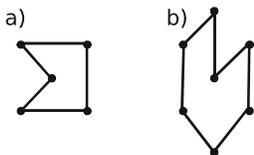

Figure 2 Example of two 2-dimensional polytopes

### B. Integer Programming

Above means that Integer Linear Programming (ILP) differs from non-integer Linear Programming (LP) not only in terms of solution values, but also in sense of space used to store exact solution for combinatorial problem. We know also that ILP can be proved to be NP-complete [10] while LP is not [9], [11]. This seems to be trade of between space used to store exact solution and time required to solve problem.

Polytopes mentioned in previous section modeled for ILP are described with limited number of variables and equations. Properly defined 0-1 polytope allows only feasible solutions (solution placed in polytopes vertex) for original problem and finding solution is equivalent to point out single polytope vertex. If problem has many solutions then integer programming solver will pick up one of them – it does not matter which one, but only one, what means that condition is satisfied – one polytope vertex was chosen. Solution is feasible, because every polytope vertexes stand for feasible solution.

Pointing out one polytope vertex requires $\Omega(\log(v))$ symbols, where $v$ represents number of vertexes. We know that there are $2^n$ vertexes for polytopes representing NP-complete problems, so number of symbols required for pointing one vertex is $\Omega(\log(2^n))=\Omega(n)$.

### C. Linear Programming

Linear programming is known be solvable in polynomial time [9], [11]. When thinking of polytope and model transition from 0-1 to linear model is like transition from set of vertexes to $k$-dimensional polyhedron. We can see that if extreme solution was found for such polyhedron it would be still on polytope lattice or facet, so it seems to have no affection on solutions ability (if we consider ability to give response to YES/NO question not ability to produce "certification" in polynomial time).

Main difference is that linear version of solution does not refer to vertexes any more. It shows one point inside polyhedron, or on facet of polyhedron.

Now referring to section II.A we can observe that this polyhedron is represented in memory as some model. Quality of model "seen" by machine is an outcome from number of factors stored in memory – greater number of factors means better quality.

Important question in discussion about using LP to solve NP-complete problem is question about minimum space required to store model of polyhedron of quality enabling to find a solution.

NP complexity class definition requires only to decision problems, what means that only answer from problem solving algorithm is YES or NO. This algorithm need not to know exact solution (certification), it only has to decide if answer to question is positive. Optimization problems such as TSP or QAP can be considered as decision problems because one may ask question "is there a Hamilton path of overall cost equal to X". Finding YES/NO answer to this question is decision NP-complete problem [10].

If then one would like to use LP solver for any NP-complete problem then at least following must occur:
1) space required to store model is polynomially bounded,
2) time required to store model is polynomially bounded,
3) time required to find solution is polynomially bounded,
4) algorithm answer YES iif there exists a solution of value X

We know that 3) is satisfied when we use LP, but other requirement cannot be satisfied. In order to prove it we shall consider example of problem which cannot possible satisfy 1),

2) and 4) in the same time. At least one of these condition has to be unsatisfied for following example.

First observation that 4) may be false refers to fact that LP model describes whole polyhedron. Asked for non-optimal solution it may answer that value X is possible (referring to some point inside polyhedron) but in fact this value may be incorrect for any of possible solutions for original problem. Of course if it was only error method generates it might have been forgiven. We have to show that method allows solutions better then optimal for original NP-complete problem.

As an example we shall consider 2 dimensional polytope containing $2^n$ vertexes. Let us assume that some of vertexes lay in path similar to arcs shape and there are $O(2^n)$ such vertexes.

Let us then consider possible target functions $f(x)=a*x+b$, for $a$ varying from 0 to $-\infty$. We will then have situation as on fig. 3.

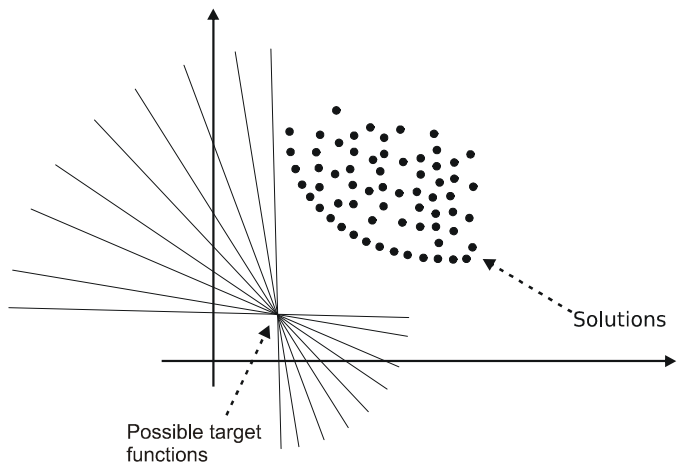

Figure 3 Solutions and possible target functions

Target functions in this example vary from almost vertical to horizontal line. In certain instance target function is picked up and then it is "moved" towards solutions until it intersects model.

Now we will show why 1), 2) and 4) cannot be true in the same time. Let us assume that 2) is true. This means that model in memory is built without checking every possible vertex (if such checking was done then 2) would be false). Model is then stored as polynomial number of equations so 1) is satisfied as well, but we do not have equation for every pair of vertexes (we said that number of vertexes on outline path is $O(2^n)$, so *lowest possible* number of pairs is also $O(2^n)$, storing them would not fir in polynomially bounded space). If we try to find solution we will find incorrect answer.

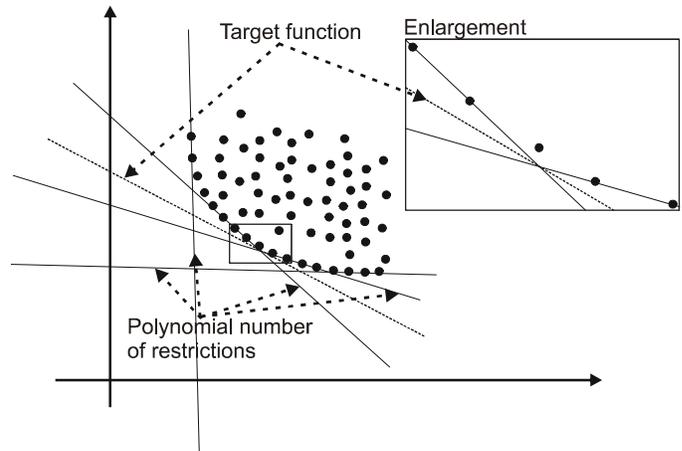

Figure 4 Limited numbers of line restrictions and target function

We can see that if number of lines is polynomial there can be defined target function equal to missing restriction (crossing omitted pair) which allows to prepare counter example returning incorrect answer to problem question. There are $O(2^n)$ different pairs what means that every model storing information about polytope in polynomially bounded space has to omit some of them.

Only possibility to build model for presented example is to start from target function definition and build whole model in such way that it would be very accurate in neighborhood of target function possible intersection. In this case for two different target functions models stored in memory would differ.

If we consider that: computer cannot store all vertexes in polynomially bounded space, so there is not known any order of these vertexes (if they were to be ordered then they must have been checked, what would take $O(2^n)$ time). Picking up neighborhood consisting of polynomial number of points (tight neighborhood) from $O(2^n)$ points seems to be at least as hard as Hitting Set Problem or Set Covering Problem described in [10]. This means that if 1) and 4) are to be true, then 2) is not.

In summary of this section we need to notice that after transition from ILP to LP number of factors stored for polytope representation must be $O(v)=O(2^n)$ for polytopes representing NP-complete problems [13], or time required to build model is equal to time required to solve another NP-complete problem, or the solver will produce incorrect answers.

IV. EXAMPLE FROM ONE OF SOLUTION ATTEMPTS

In [4] author shows solution for TSP problem introducing model containing $O(n^9)$ variables and $O(n^7)$ equations. Exactly the same model is presented in [5] and similar method is presented in [7]. Model is build up with no respect for target function (change of target function does not cause any changes in model).

This model can be proved to be incorrect for $n=32$ nodes what was actually done in [8]. This counter example was build to be as small as possible (for 32 nodes there are almost $10^6$ equations containing non-zero variables) and it used limitation of perspective in scope of each variable. Each variable represent flow on one arc, pair of arcs or triple arcs. Making instance of problem where for any selected pair there was solution fitting in model but incorrect as TSP solution was main target of this work.

Idea of counter example uses existence of "valleys" where all cities are placed and "mountains" with huge cost of crossing. There are 4 valleys in counter example for implementation presented in that article and 10 in extended version of this example for any implementation basing on same idea. This means that minimal optimal solution must consist of 4 or 10 mountains crossings (cost of travels within valley is negligible). If we assign "flows" to arcs as shown on fig. 5 for example with 10 valleys, then each 1/3 of flow consists of 9 mountains crossing, so overall cost is 9 (while optimal for TSP has to be 10). Solution found with respect to all restrictions defined for model is incorrect.

Exactly same considerations apply to QAP from [5] and TSP form [7].

In previous section we have shown that such approach may be correct iif there would be O($n!$) equations (representing TSP polytope consists of $n!$ vertexes) or model would be dependant of target function with time required to build it equal to time required for solving other NP-complete problems.

## V. Conclusion

In preceding text we have shown the difference between object and its representation. In addition difference between representation of single and set of solutions was discussed. Methods operating on single solution at time (ILP) can use O($\log(v)$) symbols and restrictions, where $v$ represents number of vertexes, what for NP-complete problems means O($n$) symbols and O($n$) restrictions. But solving ILP is as hard as solving any other NP-complete problem.

We also showed difference between ILP and LP, showing that for polytopes consisting of O($2^n$) points representation of such polytopes with less then O($2^n$) symbols or requirements used may lead to incorrect solutions even for 2 dimensions. Other way – building model to meet target function is again as hard as any other NP-complete problem.

This considerations show clearly that usage of LP for polynomial solutions of NP-complete problems simply cannot be correct, especially for large instances.

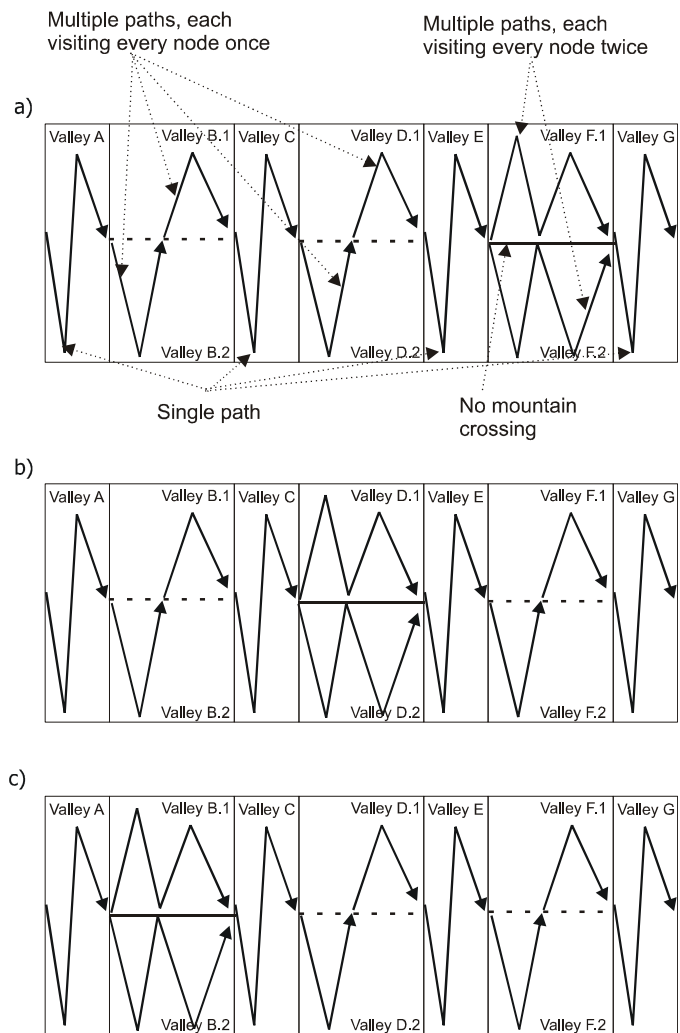

Figure 5 Idea of ten valleys for TSP counter example